# Machine learning in weekly movement prediction

HAN GUI
Department of Informatics
King's College London
Bush House, Strand Campus, 30, Aldwych, London WC2B 4BG
han.gui@kcl.ac.uk

## Abstract

To predict the future movements of stock markets, numerous studies concentrate on daily data and employ various machine learning (ML) models as benchmarks that often vary and lack standardization across different research works. This paper tries to solve the problem from a fresh standpoint by aiming to predict the weekly movements, and introducing a novel benchmark of random traders. This benchmark is independent of any ML model, thus making it more objective and potentially serving as a commonly recognized standard. During training process, apart from the basic features such as technical indicators, scaling laws and directional changes are introduced as additional features, furthermore, the training datasets are also adjusted by assigning varying weights to different samples, the weighting approach allows the models to emphasize specific samples. On back-testing, several trained models show good performance, with the multi-layer perception (MLP) demonstrating stability and robustness across extensive and comprehensive data that include upward, downward and cyclic trends. The unique perspective of this work that focuses on weekly movements, incorporates new features and creates an objective benchmark, contributes to the existing literature on stock market prediction.

## 1. Introduction

Accurate predictions of stock markets movements are crucial for investors and traders, trading strategies can be designed on the predictions to mitigate potential risks and make consistent profits. With the development of information technology, researchers and investors are increasingly utilizing machine learning techniques to forecast the future trends of financial markets, as machine learning algorithms have the potentials to analyze vast amounts of data, identify complex patterns, and make predictions based on those patterns. However, the prediction tasks are still very challenging, as financial markets are dynamic, noisy and volatile.

**EMH & RWH**

Efficient market hypothesis (EMH) and random walk hypothesis (RWH) state that prices of financial instruments already reflect all available information, and financial markets follow a random and unpredictable pattern, it is not possible to consistently beat the market with more than 50% accuracy or above-average profits [1]. While the two hypotheses are theoretical frameworks and not universally accepted. A considerable amount of research suggests that certain markets, particularly emerging markets, may not be fully efficient or well-organized. As a result, there is a possibility that predicting future stock prices and returns could yield better outcomes than random selection. Moreover, when examining the stock market through the viewpoint of behavioral economics and the socioeconomic theory, the market is predictable to some degree [3]. Namely, it is possible to forecast the market to a certain extent.

**Fundamental & technical analysis**

Generally, there are two primary methods of analyzing financial markets: fundamental analysis and technical analysis. Fundamental analysis involves the evaluations of various factors such as the intrinsic value of stocks, the performance of the industry and the economy, and the political climate, etc. Technical analysis use indicators and models derived from historical data to identify patterns and trends, and assumes the patterns and trends can provide insights into future price movements. The most commonly used technical indicators for stock market prediction include SMA (Simple Moving Average), EMA (Exponential Moving Average), MACD (Moving Average Convergence Divergence), and RSI (Relative Strength Index) [3]. Traditional methods like ARIMA (Auto-Regressive Integrated Moving Average) and GARCH (Generalized Auto-Regressive Conditional Heteroskedasticity) have been widely used in time series analysis, but machine learning models have gained more popularity in recent years, and the most frequently used machine learning models for stock market prediction are SVM (Support Vector Machine) and ANN (Artificial Neural Network) [2], [3], [5].

**Classification & regression**

Stock market prediction can be mainly categorized into two types based on the prediction target [5]:
  (1) Classification-based Prediction: it aims to determine the direction of the stock market movement, whether it will go up or go down, which can be considered as classification problem.
  (2) Regression-based Prediction: it aims to forecast the actual numerical value of a stock or index such as the closing price, which can be treated as a regression problem.

**Data & market**

Researchers often focus on specific markets, such as the developed markets in USA and Europe, the emerging markets in Asian or Middle East. And the data used in research on stock market prediction can indeed be categorized into different terms:
  (1) short term: it includes data at high frequencies, such as minutely, hourly, and

        daily intervals.
   (2) middle term: it refers to data at weekly and monthly intervals.
   (3) long term: it encompasses yearly intervals.

Kumbure et al. [5] reviewed 138 journal articles and found that the majority of studies concentrated on daily predictions, additionally, the USA market received the most attention. After examining 124 papers, Jiang [4] confirmed that the majority of studies (105 out of 124) focused on daily predictions. Based on the review of 57 texts, Henrique et al. [2] also observed a predominance of studies using data from the USA market.

**Feature engineering & model selection**

Market prediction typically involves two main components: feature engineering and model selection. Feature engineering is the process of extracting meaningful variables from the available data, which can capture relevant information about the market dynamics. Once the features are engineered, the next step is to select an appropriate model that can effectively learn from the data and make predictions. There are various machine learning algorithms available, the most frequently used machine learning models for stock market prediction are SVM and ANN [2], [3], [5], other models such as RF (random forest), LR (logistic regression) and NB (Naïve Bayes), besides the SVM and ANN mentioned above. While the selection of a model depends on various factors, including the specific nature of the prediction task, the size and quality of the available data, the desired level of interpretability, and the computational complexity involved.

**Measurement & performance**

To evaluate the performance of machine learning models in stock markets predictions, there are different types of measurements used. These measurements can be categorized into two main groups: technical and financial. Technical measurement is to measure the overall correctness of the predictions, such as accuracy / hit ratio, precision and recall, F-score, and Matthews Correlation Coefficient (MCC) [12]. Financial measurement is to measure returns generated by the trading strategy designed on the predictions, as the ultimate goal is to generate profits and the profitability directly reflects the success [9], [18], [21]. More than 87% of the papers reviewed reported that their model beat their benchmark model [3], while their benchmark models are not the same. Performance is indeed relative, and the creation of a standardized benchmark is essential to enable meaningful comparisons across different models. Technical measurement is direct and standard such as the accuracy, precision and recall, they are calculated by standard formulas, while financial measurement needs to take into account the trading strategy, and the baseline of profitability such as buy-and-hold (buy on the first day and sell at the end) or movement-trade (if today's index is higher than yesterday then buy, otherwise sell) are too weak and not very practical [9], [18], [21].

This work focuses on classification-based stock market prediction, it can also be generalized into other time series prediction problems in financial markets such as cryptocurrency, commodity, bond and exchange rate. This study introduces novel ways of feature engineering and data augmentation, and trains several models on extensive and diverse datasets, then bring up a new benchmark of random traders, which is independent of any ML model and thus more objective measurement for the performances of models. The main contributions of this work:
- a) provides a clear message that utilizing machine learning techniques for weekly movement prediction in stock markets can be feasible and profitable
- b) offers a general and transparent workflow and overview of the process, making it accessible and easy to reproduce for newcomers in this field
- c) enhances the credibility and persuasiveness by using extensive data that covers different trends
- d) introduces novel features that are likely to improve the predictive capabilities of the models
- e) presents original approaches to data augmentation
- f) establishes an independent and objective benchmark of random traders

The remainder of this paper is organized into following sections: section 2 reviews the existing literature on classification-based stock market prediction, section 3 describes the methodology adopted for stock market prediction, section 4 discusses the results of classification-based stock market prediction, and section 5 summarizes the key findings and contributions of this work.

## 2. Literature review

Stock market prediction by machine learning as a classification problem has been researched for a long time, in this section some of the highly cited papers are listed and discussed from several perspectives.

**Features engineering**

In order to capture a broader range of market dynamics and improve prediction accuracy, features can be extracted from various data sources. Technical indicators calculated according to historical data (prices and volumes) are often included in studies **[7], [10], [13], [15], [17], [18], [20], [22], [23], [25]**, such as SMA, MACD, and RSI. Macroeconomic data such as interest rate, exchange rates and consumer price index, which reflect the economic circumstances of a particular country, region or sector, are also used sometimes **[6], [9], [15]**. Text data from social media or financial news can provide insights into market sentiment and potentially assist in predicting stock market movements, and natural language processing techniques, such as bag-of-words, TF-IDF (Term Frequency-Inverse Document Frequency) and word embeddings, are commonly applied to transform textual data into numerical representations, namely, to extract the sentiment scores **[8], [11], [12], [14], [16], [19], [21], [24]**.

Normalization is a common technique used by researchers to ensure that the values of input features do not disproportionately influence the prediction model, which can mitigate the impact of varying scales and improve the stability **[7], [9], [13], [14], [15], [22], [25]**.

Furthermore, some researchers employ discretization or binarization techniques, aim to simplify the representation of technical indicators and represent the trend by converting continuous parameters to discrete value such as '+1' or '-1' based on the property of each technical indicator, which improves the overall forecasting performance **[13], [25], [22]**.

There are a few papers that use feature selection and weighting to enhance performance, for example, cluster features into different groups and select the most informative one from each group by ranking, or use a SVM model to assign different weights for different features **[10], [17]**.

Using data from overseas stock markets and other financial markets as input features can also be a valuable approach, especially when there is a strong temporal correlation between markets. For example, predict Japanese market by introducing the latest USA market data as USA is the largest export target for Japan, forecast USA markets by incorporating the latest data from European and Asian markets as their movements already account for possible market sentiment on latest global dynamics **[6], [9]**.

**Models**
SVM and ANN are widely employed in this field, SVM can utilize different kernel functions, such as linear, polynomial, and RBF (Radial Basis Function), ANN often refer to MLP (Multi-Layer Perceptron) or LSTM (Long Short-Term Memory). Some researchers compared several models and generally found that ANN has better performance **[7], [12], [13], [15], [18], [22], [23], [24], [25]**, the reason may be the advancements of deep learning in recent years.

Regarding ensemble methods, they are also frequently utilized in stock market prediction studies, as it can improve predictive performance compared to individual learner, and easier to implement than single complex model. Ensemble methods, such as RF (Random Forest), AB (AdaBoost), and GB (Gradient Boosting), are commonly applied in stock market prediction tasks, and RF is particularly popular among ensemble methods **[13], [16], [18], [20], [22], [23], [24], [25]**.

Other simpler models such as LR (Logistic Regression), KNN (K-Nearest Neighbor), NB (Naïve Bayes) have advantages like interpretability, computational efficiency, and ease of implementation, still receive attention as the baseline in stock market prediction research **[17], [21], [22], [24], [25]**.

Two-stage models, where one model is used for feature selection, processing or weighting, and another model is used for making predictions based on those features, are claimed to help improve prediction accuracy by a few works **[16], [17]**.

**Training & testing**

For most of studies reviewed, the information about training is not given such as the training accuracy **[6], [9]. [10], [12], [14], [16], [17], [18], [20], [21], [22], [23], [25]**, which may inevitably raise some concerns about overfitting. In study **[7]**, the training accuracy is almost 100% while testing accuracy is much lower, which is obviously overfitting and risky for investors to put into practice. Some studies claiming 80-90% or even 90+% accuracies such as **[13], [21], [25]** may suffer from lookahead bias, that is, the training data overlaps with the testing data in terms of time range, in other words, use the future information to predict the past especially when shuffling the datasets before splitting, which will lead to artificially inflated accuracy results. The graph below shows an example:

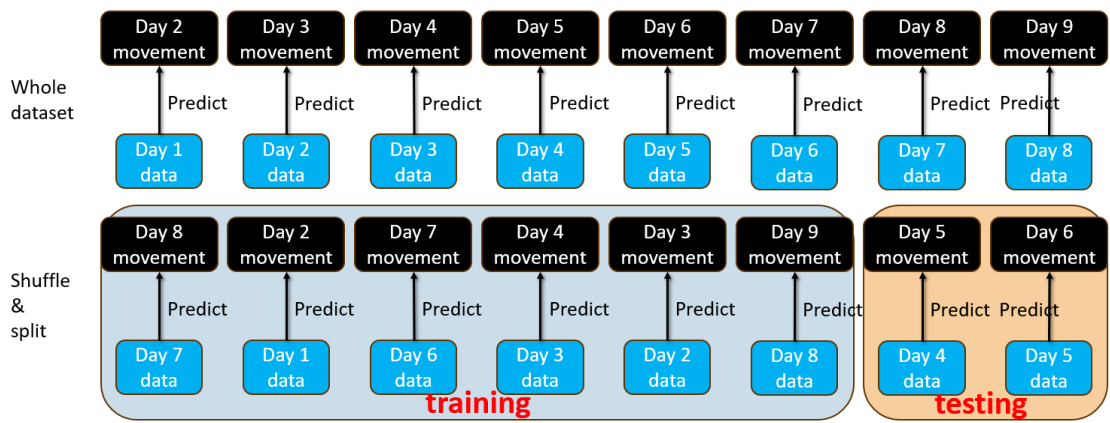

(figure 1: the example of look-ahead bias, the fundamental idea is to use data from day $T$ to predict $T+1$, while in this particular example, after shuffle and split the datasets, the training set includes future data (day 7, 8 and 9), while the testing data consists of past observations. This means that the future data is being used to predict the past, which introduces a lookahead bias)

## 3. Methodology

### Datasets

Weekly data (5 days)

This work is going to predict weekly movements of stocks markets, that is, use current week information to predict next week movement. Compared with daily predictions that provide forecasts for each trading day, capturing short-term market movements and fluctuations, weekly predictions have several advantages as follows:

(1) offering a broader perspective and insights into medium-term trends and patterns
(2) smoothing out some of the day-to-day volatility
(3) more accessible and less computationally intensive
(4) suitable for investors with longer investment horizons or those who prefer a less frequent trading approach or more passive investment approach, such as retail investors

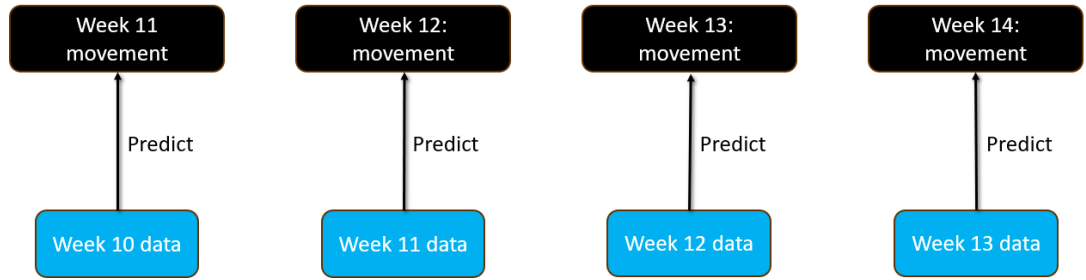

(figure 2: the example of weekly prediction, use data of week $T$ to predict the movement of week $T+1$)

Indexes / stocks

Most of the studies focus on a single market **[4]**, with many of them exclusively applying machine learning models to a specific index or stock, such as S&P 500 Index **[24]**, SSE 50 Index **[21]**, NIKKEI 225 index **[6]**, Istanbul Stock Exchange (ISE) National 100 Index **[7]**, IBovespa index **[18]**, Microsoft **[16]**, Apple **[19]** and Bitcoin **[23], [25]**.   While this work tests machine learning models on much more extensive data, covering the main American indexes (Dow Jones, S&P 500 and Nasdaq), and top American stocks (Microsoft, Apple, Nvidia, Alphabet (Google), Amazon, Meta (Facebook)), and the results will be more convincing as a wider range of indexes and stocks can exhibit more diverse market dynamics and provide a more robust assessment for machine learning models.

For the two graphs below, the first chart (figure 3) displays the weekly price / point changes of index / stock. Meanwhile, the second graph (figure 4) illustrates the relative price / point of each index or stock in comparison to its starting time, for instance, if the value is 100 in week one and 110 in week two, they are converted to 1 and 1.1, respectively. This relative representation reflects the overall trend of each index / stock, encompassing upward, downward and cyclic trends, which shows the datasets used in this work are extensive and comprehensive.

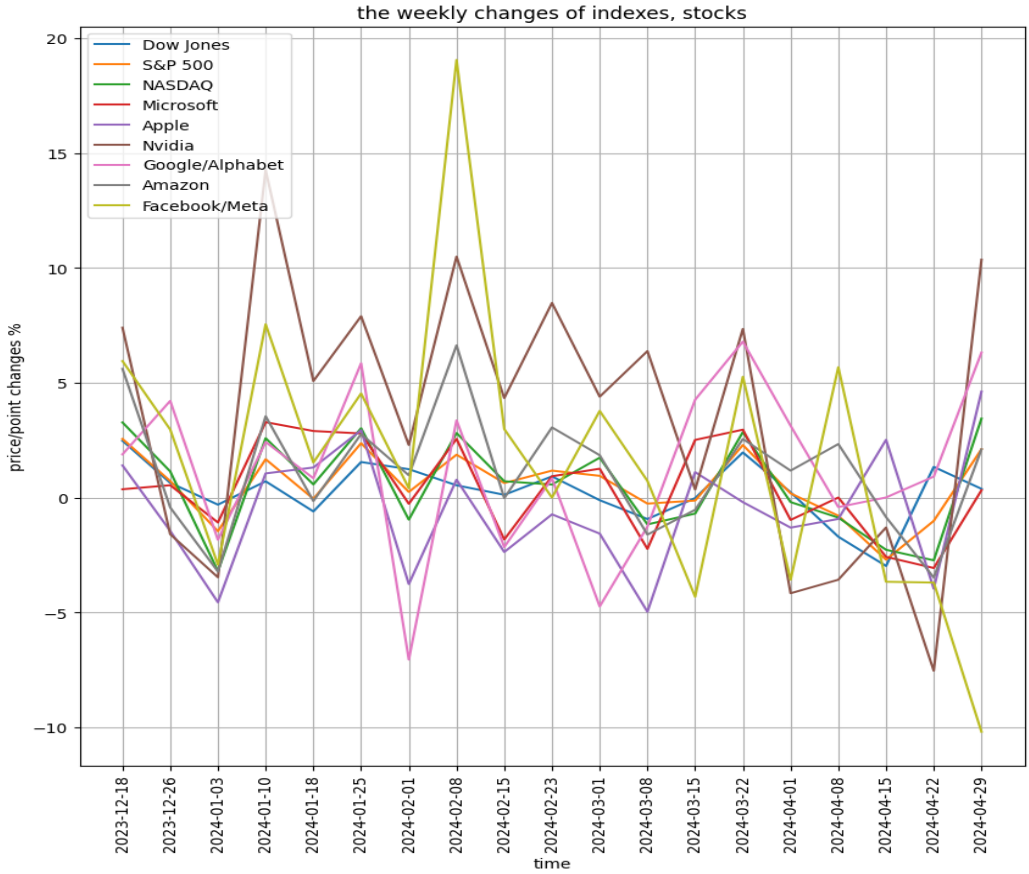

(figure 3: the weekly changes of the main American indexes and top stocks)

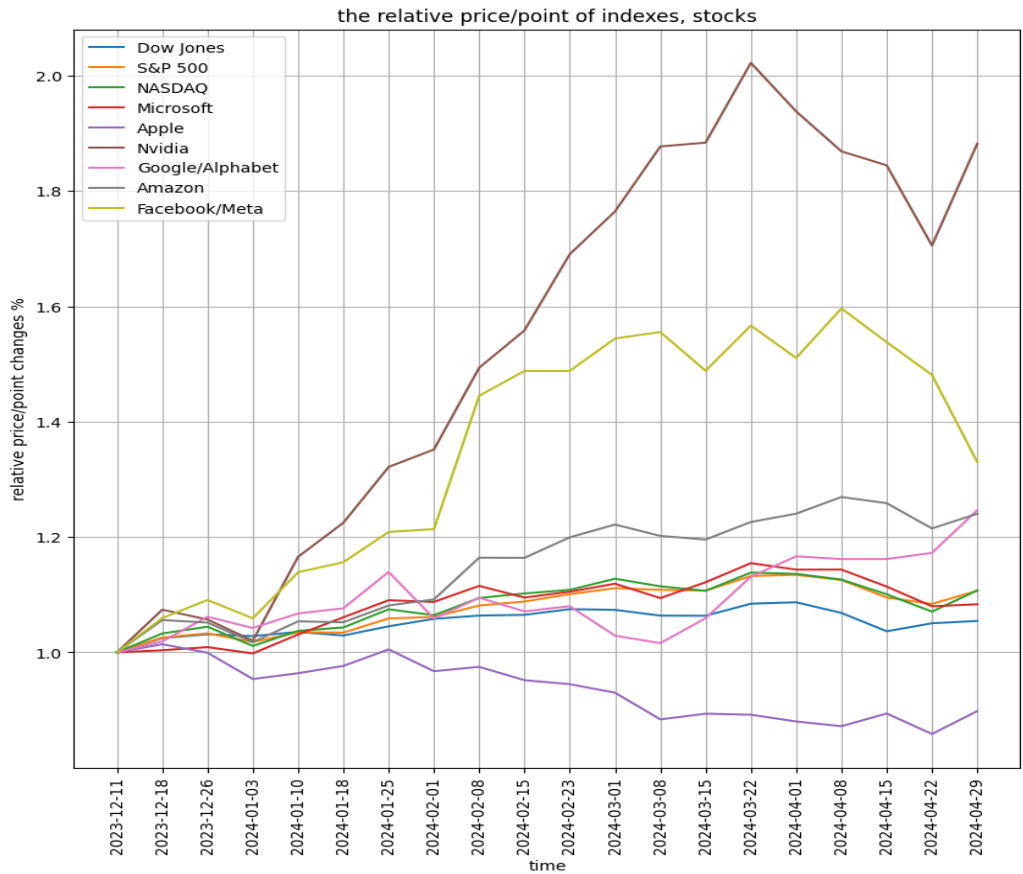

(figure 4: relative price / point of the main American indexes and top stocks)

Data augmentation

This study uses Yahoo finance to collect data of the main American indexes and top American stocks from 2021-01-01 to 2024-05-01, each dataset is divided as two parts: training (in-sample) and testing (out-of-sample) datasets, and the size of testing dataset is 20 that is convenient for the evaluation phase.

Data augmentation widely used for image classification and object detection tasks, and proved to effectively enhance the classification and detection performance. After examining 124 papers, it is found very few works explore the usage of data augmentation **[4]**. In this work, training dataset is augmented by giving more duplicates to samples with bigger changes, here the duplicates do not imply over-fitting, they give that sample more weights in the training, and the number of duplicates depends on the return change of this sample. The logic is: from the perspective of profitability, the prediction of each sample should not be equally weighted, for example, the sample with 10% rise or fall has multiple times of influence for the return of investment in comparison to the sample with 1% rise or fall, therefore, the sample with bigger change should be weighted more. This will let the models focus more on the samples with jumps. The duplicate number of each sample is computed by the following formula (the **return change** is weekly growth in this work):

$$Duplicate\ number\ =\ Max(\ log(|return\ change|), 1\ )$$

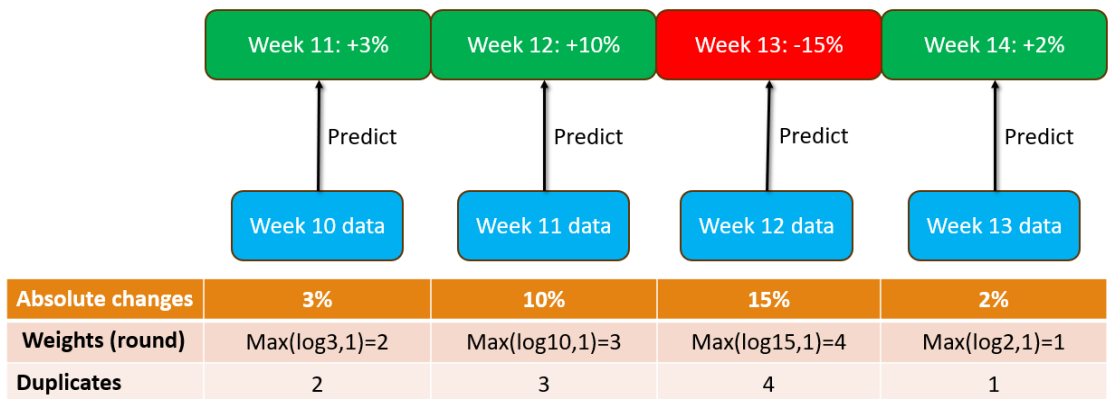

(figure 5: the example of the calculation of duplicate number)

**Feature engineering**

The features used in this work mainly include technical indicators, scaling laws and directional changes that are introduced from high-frequency finance.

Technical indicators

Technical indicators exhibit varying effectiveness depending on the market conditions. Some indicators perform well in trending markets, while others are more suitable for non-trending or cyclical markets. In this study, five specific technical indicators are utilized:

| Name | Formula |
| --- | --- |
| MACD | $EMA_{12} - EMA_{26}$<br>$EMA = exponential\ moving\ average$ |
| RSI | $$100 - \frac{100}{1 + \frac{Average\ gain}{Average\ loss}}$$<br>$Average\ Gain = mean\ of\ last\ 14\ days\ price\ gains$<br>$Average\ Loss = mean\ of\ last\ 14\ days\ price\ gains$ |
| Bollinger Band | $SMA(TP, n) \pm m * \sigma(TP, n)$<br>$TP = typical\ price$<br>$n = number\ of\ days$<br>$\sigma = standard\ deviation$ |
| BIAS | $(Close - N\ day\ average\ price)/N\ day\ average * 100\%$ |
| ATR | $$ATR_t = \frac{ATR_{t-1} * (n-1) + TR_t}{n}$$<br>$n = number\ of\ days$<br>$TR = max[(high - low), abs(high - previous\ close), abs(low - previous\ close)]$<br>$$ATR_1 = \frac{1}{n}\sum_{i=1}^{n} TR_i$$ |

Scaling laws

In financial markets, the scaling laws imply certain patterns and relationships between time and some variables, such as price changes and volume changes.

The statistics shown below suggests that, as time progresses, the cumulative sum of the absolute daily price movements (up or down) increases linearly, and the cumulative sum of the daily volume changes (positive or negative) also follows a linear relationship, the observations imply that, on average, the overall price and volume volatility grow steadily over time, as shown by the charts below (figure 6-11).

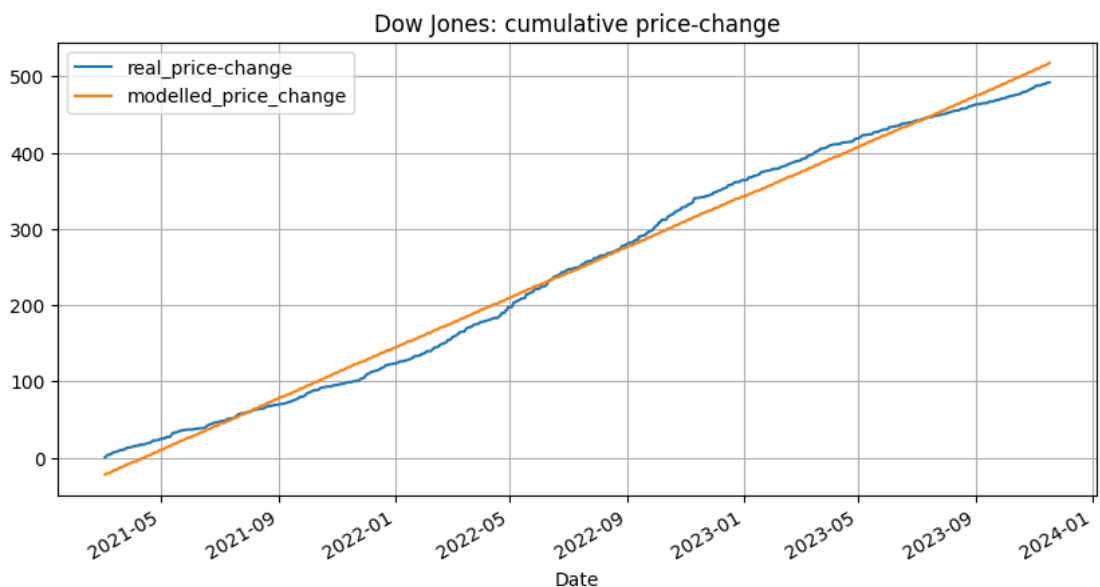

(figure 6: scaling law for Dow Jones Index over price)

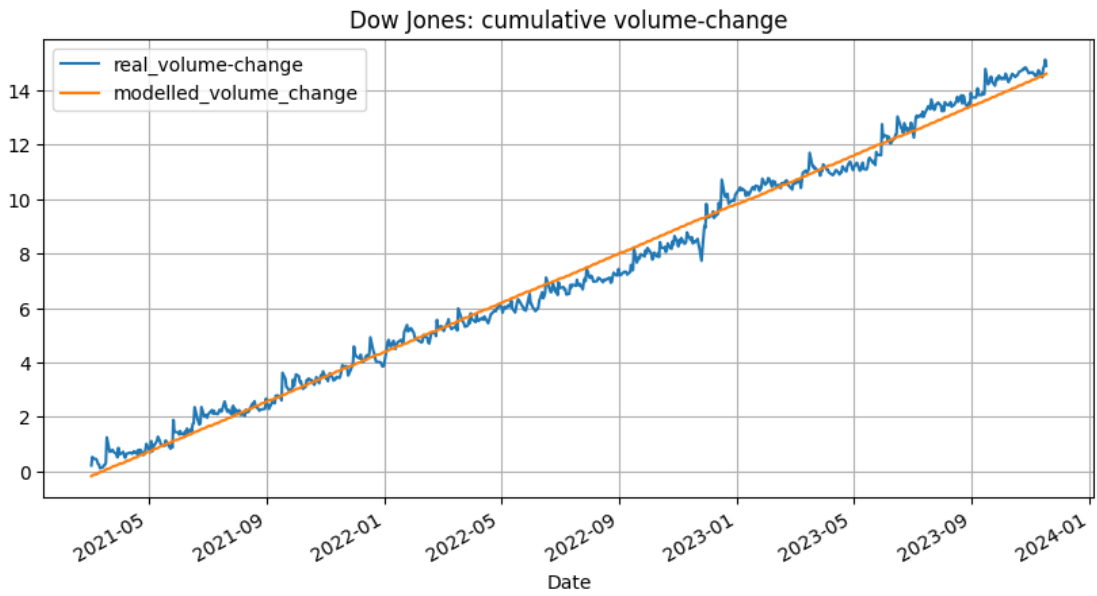

(figure 7: scaling law for Dow Jones Index over volume)

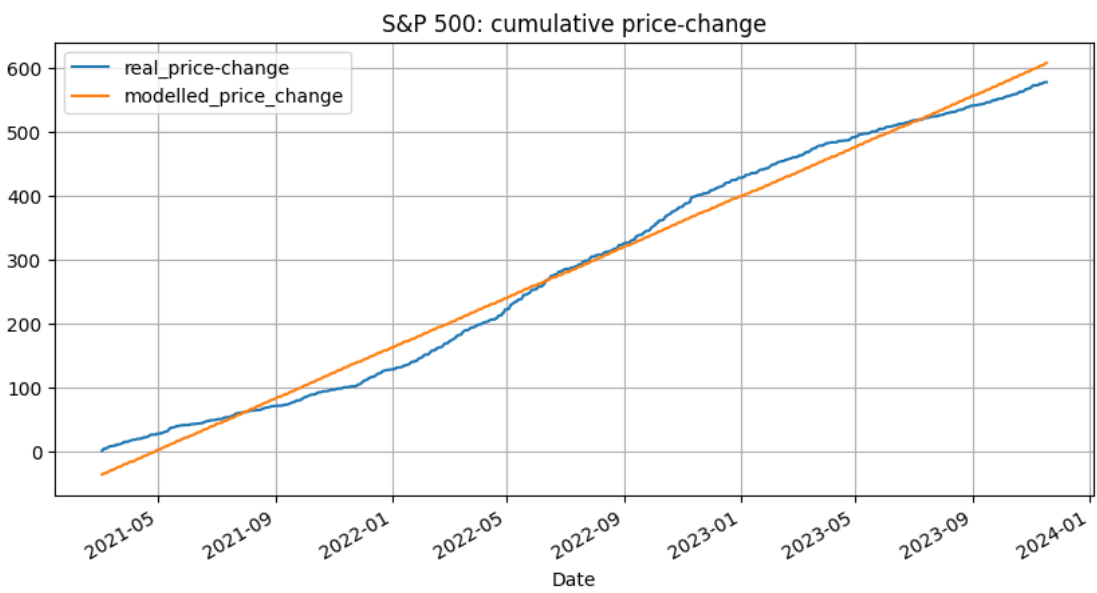

(figure 8: scaling law for S&P Index over price)

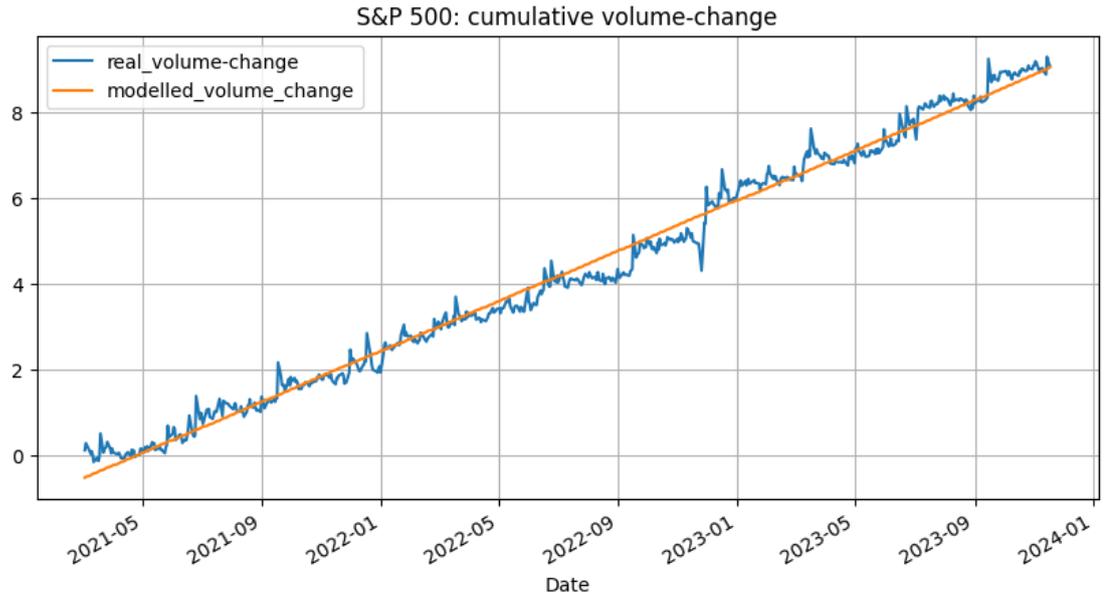

(figure 9: scaling law for S&P Index over volume)

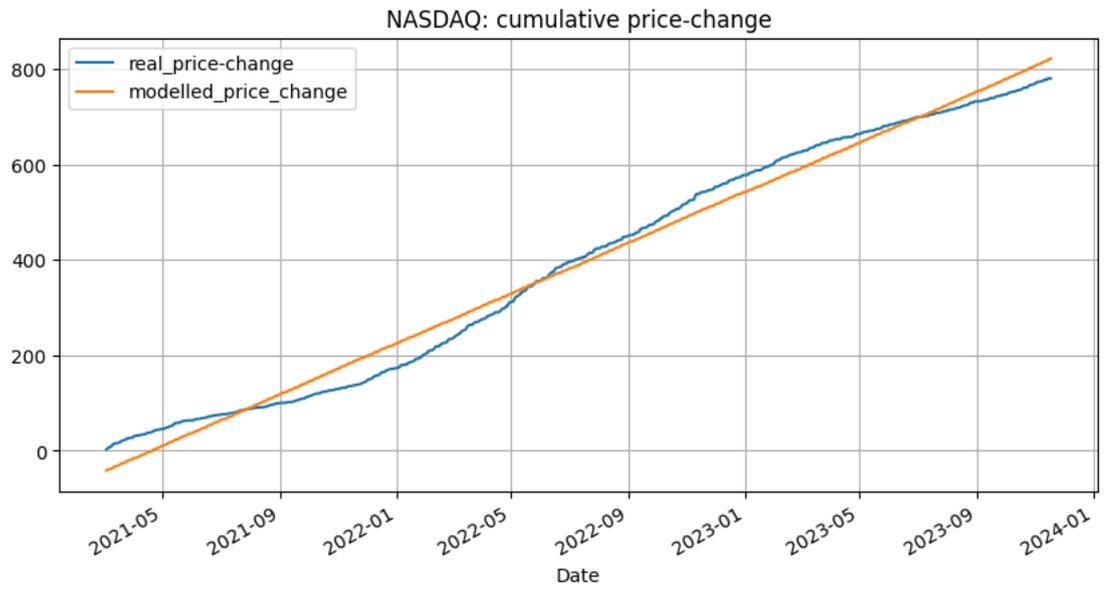

(figure 10: scaling law for NASDAQ Index over price)

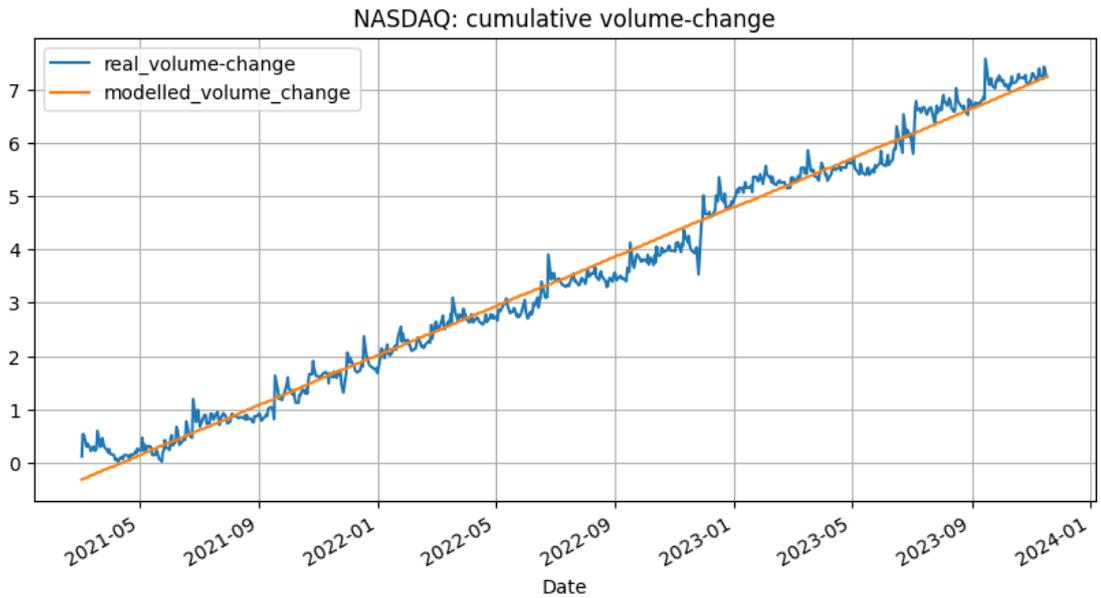

(figure 11: scaling law for NASDAQ Index over volume)

For directional changes **[26]**,
Directional changes in high-frequency finance refer to shifts in the price direction of an asset over time intervals, it dissects a data series into a set of up and down trends that alternate with each other. A price that reverses a trend is called Directional change (DC), and a price move that extends a trend is called Overshoot (OS). An upward directional change event is marked as +1 and downward directional change as -1, an upward overshoot event marked as +2 and downward overshoot as -2 (see figure 12).

And the cumulative number of DC also follow the scaling laws, +1 DC is almost symmetric to the -1 DC, namely, they have the same frequency to appear. Their sum is almost to 0 over time, which can be used to predict future trends, +1 DC will be followed by -1 DC by the same frequency (see figure 13-15).

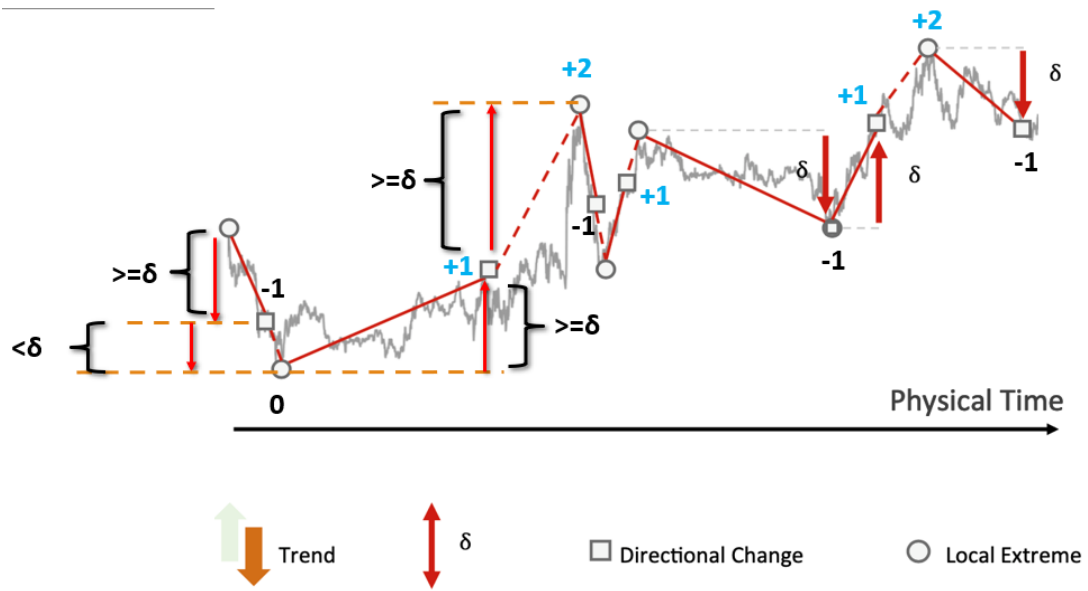

(figure 12: the illustration of directional changes. The δ is self-defined threshold such as 3% or 5%, etc., when the price change is greater than δ, then certain signal such as OS or DC is recorded.)

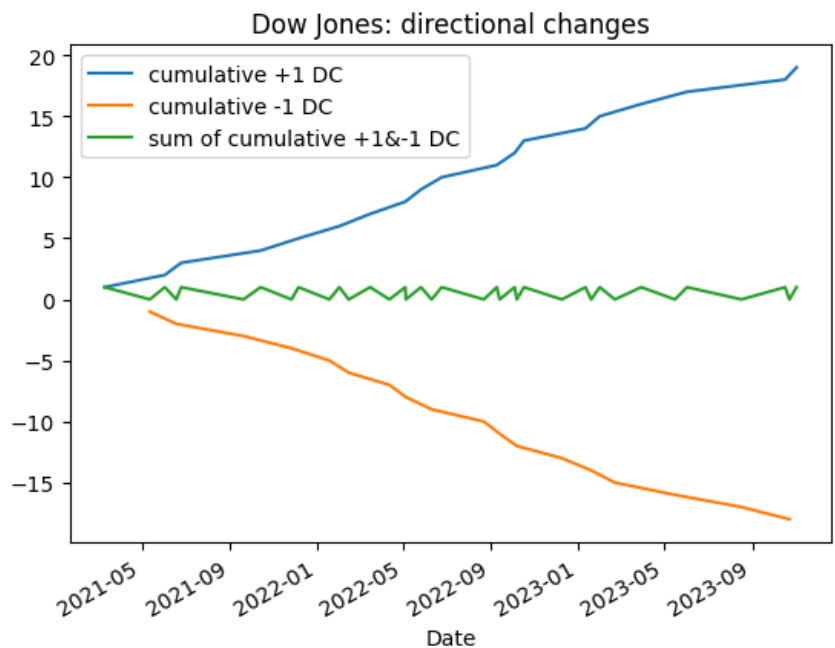

(figure 13: cumulative numbers of +1/-1DC with time for Dow Jones)

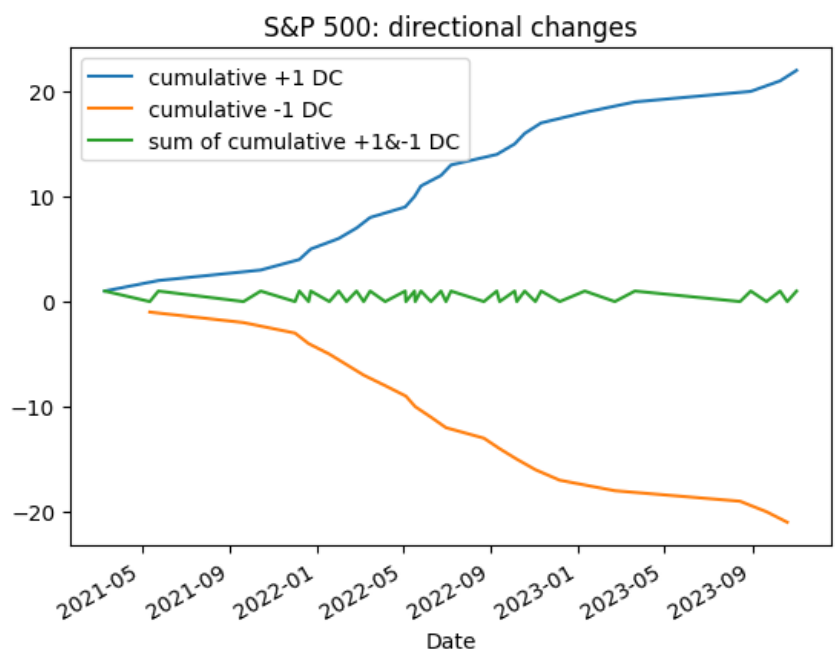

(figure 14: cumulative numbers of +1/-1DC with time for S&P 500)

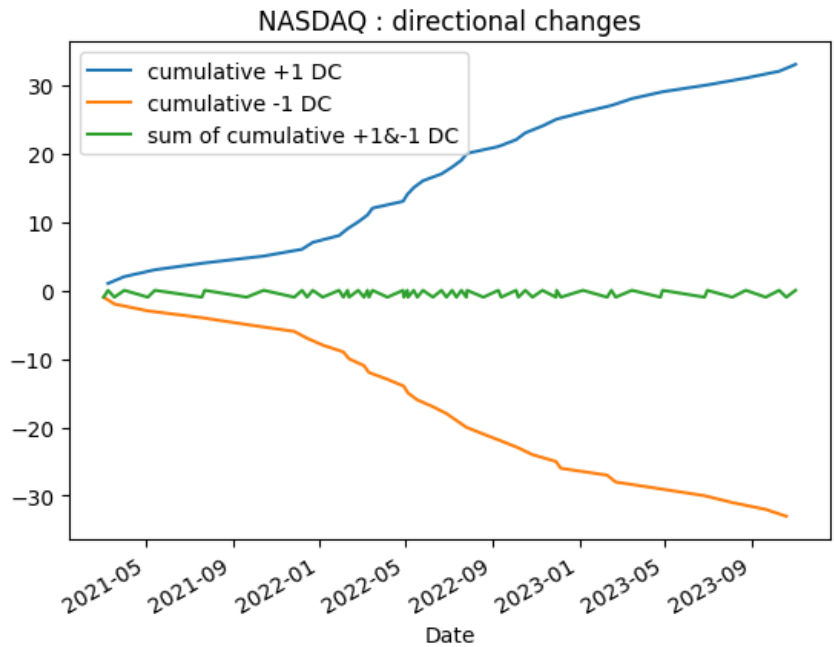

(figure 15: cumulative numbers of +1/-1DC with time for NASDAQ)

Those features are then be discretized prior to applying PCA (principal component analysis) for compression. PCA is a statistical technique for dimensionality reduction by transforming a large set of variables into a smaller set while still retaining most of the variability in the original dataset. And the basic flowchart is depicted in the graph below:

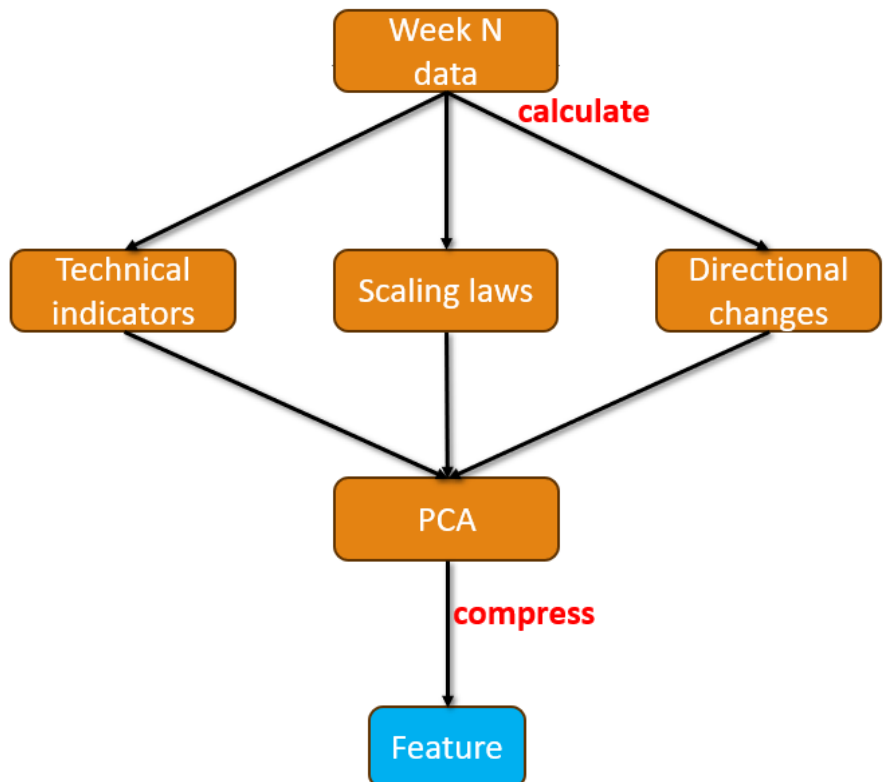

(figure 16: feature engineering)

## Models

An extensive range of machine learning models are included in this study, and the hyperparameters for each model are adjusted by numerous experimental works for optimal results. Here's a brief overview of the models: K-Nearest Neighbor (KNN), Logistic Regression (LR), Random Forest (RF), Adaptive Boosting (AB), Gradient Boosting (GB), Bagging (Bag), Support Vector Machine (SVM) and Multi-Layer Perceptron (MLP).

| Models | Details |
|---|---|
| KNN | KNN is a non-parametric classification algorithm that assigns a class label to a data point based on the majority class among its k nearest neighbors in the feature space. |
| LR | LR is a linear classification algorithm that models the relationship between the input variables and the probability of belonging to a specific class. It utilizes a logistic (or sigmoid) function to map the input to a probability value and widely used for binary classification tasks |
| RF | RF is an ensemble learning method that combines multiple decision trees to make predictions. Each tree is trained on a random subset of the data and features, and the final prediction is determined by aggregating the predictions of individual trees. |
| AB | AB is an ensemble learning algorithm that iteratively trains weak classifiers and assigns higher weights to misclassified samples. It combines the predictions of multiple weak classifiers to make a final prediction, particularly effective in handling imbalanced datasets. |
| GB | GB is another ensemble learning technique that builds an ensemble of weak prediction models in a stage-wise manner. It trains each model to correct the mistakes made by the previous models, gradually improving the overall performance. |
| Bag | Bagging is an ensemble technique that trains multiple independent models by employing bootstrapping to randomly select subsets of the training data, each with the same size as the original dataset. During prediction, all models make individual predictions, and the final prediction is obtained by aggregating the predictions from all the models. |
| SVM | SVM is a powerful supervised learning algorithm used for classification, finds an optimal hyperplane that maximally separates the data points of different classes by using different kernel functions to map the data into higher dimensions. |
| MLP | MLP is a type of artificial neural network with multiple layers of interconnected nodes (neurons). It can be used for both classification and regression tasks. MLPs are known for their ability to learn complex patterns and relationships in data but may require careful tuning of hyperparameters and regularization techniques to avoid overfitting. |

## Financial measurement – random traders

The prediction of model is not the end of the journey, the profitability of the trading strategies based on the prediction of model should be analyzed. The profitability of a model is to check the return of investment when trading based on predictions of the model, a simple and practical trading strategy is to buy when prediction is positive and to sell when prediction is negative [9].

The following graph shows the profitability of the trading strategy based on a perfect prediction, its accuracy is 100% and return of investment is +15.57%.

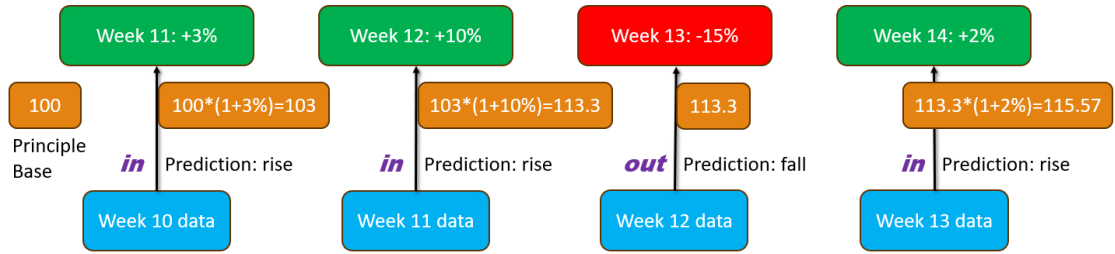

(figure 17: the example of how to calculate the profitability)

Regarding the lack of a commonly recognized benchmark for evaluating the profitability of machine learning models on stock market predictions, it can be challenging to compare the performance of different models, assess their relative effectiveness and draw meaningful conclusions without a standardized benchmark, therefore, a benchmark of random traders is created in this work:

Random traders:
  a) Random traders, each week trader can enter the market or not, if there are $N$ weeks, there will be $2^N$ cases, for example, there are about 1 million cases for 20 weeks. When the time range increases, the number of cases rise exponentially and it becomes impractical to calculate for long term. While in the context of market prediction, it is common to focus on short or medium-term predictions, this is because market conditions can change rapidly, model should be updated with the new-coming information to allow it to stay in sync with the market and adjust its predictions accordingly. Thus, it is still reasonable and feasible. For week $t$, the trader can enter the market ($in$) or do not stay in the market ($out$), the state of week $t$ can be expressed as:

$$s_t = \begin{cases} 0, & out \\ 1, & in \end{cases}$$

  b) If the trader enters the market ($in$) in week $t$, its principal will go up by weekly return when market rises or go down by weekly return when market falls. If the trader does not stay in the market ($out$), its principal will not be influenced by the market whatever the market rises or falls, the principal change for week $t$ can be expressed as:

$$C_t = s_t * R_{weekly\ return}$$

  c) Benchmark is the expected return of all random traders, namely, the average of gross returns of all cases. The return of case $i$ over $T$ weeks expresses as:

$$R^i = \prod_{t=1}^{T}(1 + C_t), case\ i = \{s_1^i, s_2^i, \ldots, s_T^i\}$$

The expected gross return of all case expresses as:

$$E = \frac{1}{N}\sum_{i=1}^{N} R^i, for\ N\ cases$$

The below graph (figure 18) gives an example for the random traders, "**in**" means have a position in the market, and "**out**" means do not stay in the market.

| | 1 | 2 | 3 | 4 | 5 | 6 | 7 | 8 | 9 | 10 | 11 | 12 | 13 | 14 | 15 | 16 |
|---|---|---|---|---|---|---|---|---|---|---|---|---|---|---|---|---|
| Week 11: +3% | in | in | in | in | out | in | in | out | in | out | out | in | out | out | out | out |
| Week 12: +10% | in | in | in | out | in | in | out | in | out | in | out | out | in | out | out | out |
| Week 13: -15% | in | in | out | in | in | out | in | in | out | out | in | out | out | in | out | out |
| Week 14: +2% | in | out | in | in | in | out | out | out | in | in | in | out | out | out | in | out |
| return | 0.98 | 0.96 | 1.16 | 0.89 | 0.95 | 1.13 | 0.88 | 0.94 | 1.05 | 1.12 | 0.87 | 1.03 | 1.1 | 0.85 | 1.02 | 1 |
| Average: 1.00 | | | | | | | | | | | | | | | | |
| accuracy | 75 | 50 | 100 | 50 | 50 | 75 | 25 | 25 | 75 | 75 | 25 | 50 | 50 | 0 | 50 | 25 |

(figure 18: the example of random traders for 4 weeks)

For the relationship between accuracy and profitability of random traders from the above example, it can be found that:

(1) Same accuracy can correspond to different profitability for individual case, even if two random traders have the same accuracy in their predictions, their actual profitability can still vary, such as cases 1, 6, 9, 10, they all have the same accuracy (75%) while different returns (0.98, 1.13, 1.05, 1.12).
(2) Same profitability can correspond to different accuracy for individual case, and it is possible for random traders with different accuracies to achieve almost the same profitability such as cases 4 and 7, they have almost the same return while different accuracies (50%, 25%).
(3) Higher accuracy leads to higher profitability when calculating the expected return of cases with the same accuracy such as the cases with 75% accuracy generally have higher returns than that with 50% accuracy.

This benchmark is independent of any model and thus objective for measuring profitability. To my knowledge, no effort has been reported in the literature about this objective and independent benchmark.

Specific Random traders: random traders with the certain accuracies
The benchmark of random traders above does not define the accuracy, for all cases of the random traders above, accuracies vary from 0% to 100%. Here, a certain accuracy can be set to make it more specific.

For example, when set the accuracy as 50%, it means half predictions are correct. For the above example, it has 4 predictions, when accuracy fixed at 50%, it means 2 of the 4 predictions are correct and 2 are incorrect. Its expected gross return is 0.99 and has the following cases (figure 19):

|   | 1 | 2 | 3 | 4 | 5 | 6 |
|---|---|---|---|---|---|---|
| Week 11: +3% | in | in | out | in | out | out |
| Week 12: +10% | in | out | in | out | in | out |
| Week 13: -15% | in | in | in | out | out | out |
| Week 14: +2% | out | in | in | out | out | in |
|   | 0.96 | 0.89 | 0.95 | 1.03 | 1.1 | 1.02 |

**Accuracy= 50%, expected: 0.99**

(figure 19: example of random traders with accuracy=50%)

For the above example, it has 4 predictions, when accuracy fixed at 75%, it means 3 of the 4 predictions are correct and 1 is incorrect. Its expected gross return is 1.07 and has the following cases (figure 20):

|   | 1 | 2 | 3 | 4 |
|---|---|---|---|---|
| Week 11: +3% | in | in | in | out |
| Week 12: +10% | in | in | out | in |
| Week 13: -15% | in | out | out | out |
| Week 14: +2% | in | out | in | in |
|   | 0.98 | 1.13 | 1.05 | 1.12 |

**Accuracy= 75%, expected: 1.07**

(figure 20: example of random traders with accuracy=75%)

The graph below (figure 21) shows the expected net return ($gross\ return - 1$) of random traders with certain accuracies for the main American indexes and top 15 stocks over 20 weeks from December 2023 to April 2024, The term 'any' encompasses all possible cases with accuracies from 0% to 100%, and its performance is almost the same as random traders whose accuracy is set at 50%, and the accuracy=50% represents the average return across all cases with accuracy equal to 50%. It shows that accuracy=40% can achieve positive profit for all indexes and most of stocks, while accuracy=60% can guarantee a profit for all indexes and stocks. And the profitability of random traders with accuracy=50% can be considered as the market performance, if the trading strategy based on model prediction consistently achieves a higher profit, it suggests beating the market and challenging the EMH and RWH.

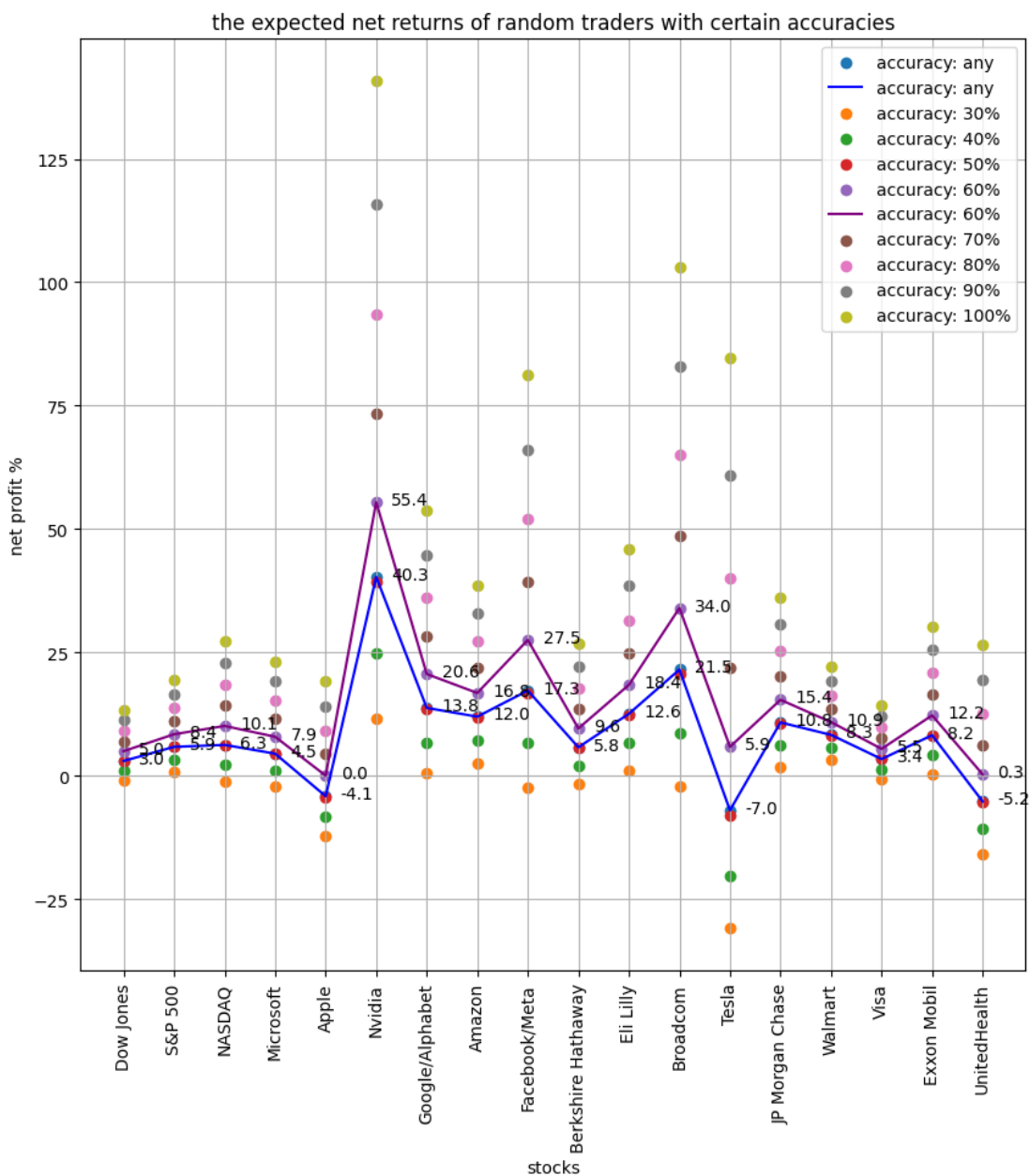

(figure 21: profitability of random traders with certain accuracies over indexes / stocks)

## 4. Results and Discussion

(1) The chart below illustrates the financial performance of ML models on prominent American indexes, i.e., Dow Jones, S&P 500 and NASDAQ. The benchmark of random traders has a positive net return across all three indexes.
    a) Overall, all models achieve more profits than the benchmark, this could be attributed to the fact that the indexes behave like a portfolio, they are stable and in an upward trend during the period.
    b) For S&P500, LR under-performs than the benchmark
    c) RF has best performance on average, Bagging shows the most stable performance (about 10% on each index), SVM and MLP also have consistently better performance than the benchmark.

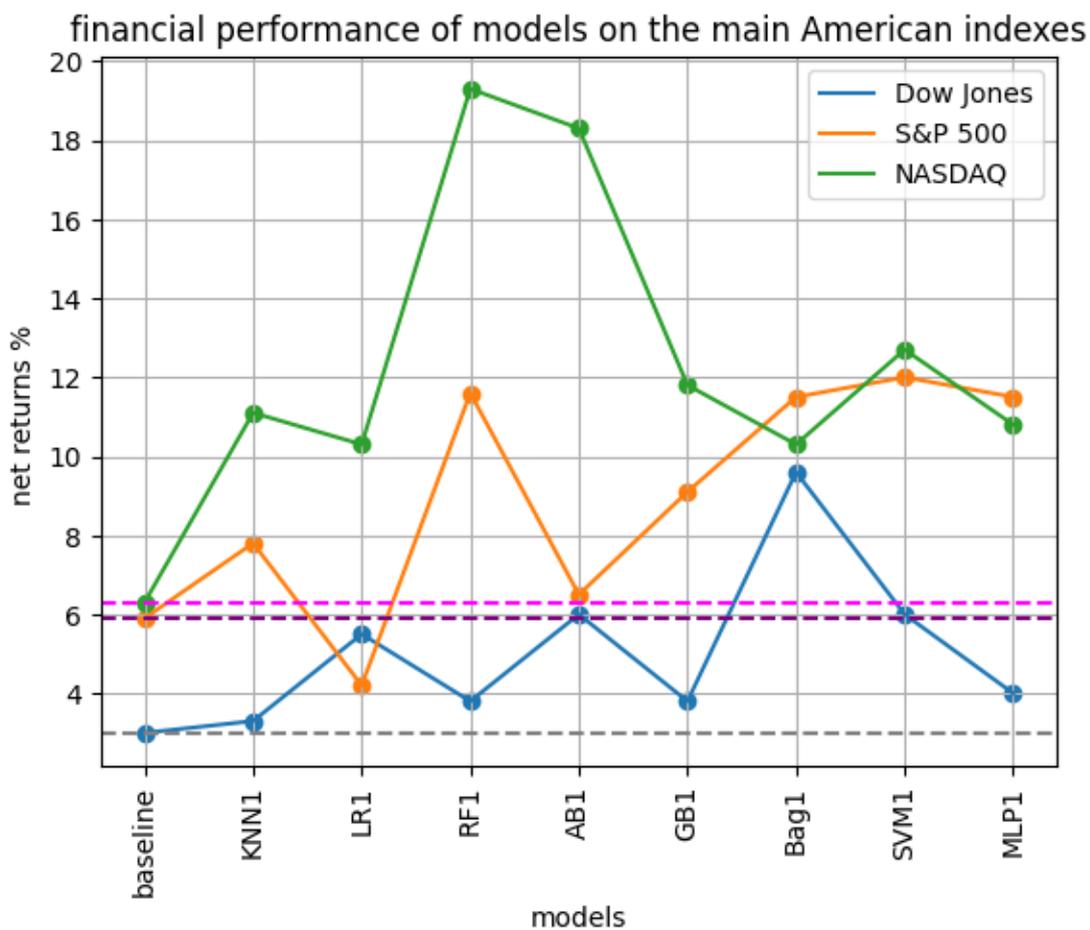

(figure 22: financial performance of different models on the main American indexes)

(2) The following charts depict the technical performance of ML models on the three primary American indexes.
    a) All models exhibit more than 50% accuracy on all three indexes, except for the underfitting observed in LR for S&P 500, where the testing accuracy is below 50%.
    b) Notably, the training accuracies of GB and SVM approach 100% for all three indexes, while their testing accuracies are about 60%, indicating potential overfitting and posing a risk for practical use. RF and KNN also display signs of

overfitting, as there is a gap of more than 20% between training and testing accuracies for Dow Jones and NASDAQ indexes, respectively.
c) On the other hand, AB, Bagging, and MLP models perform better and demonstrate greater stability, with training and testing accuracies being closer.
d) Additionally, a few observations of testing accuracy a little higher than the training accuracy, may be due to data augmentation techniques employed.

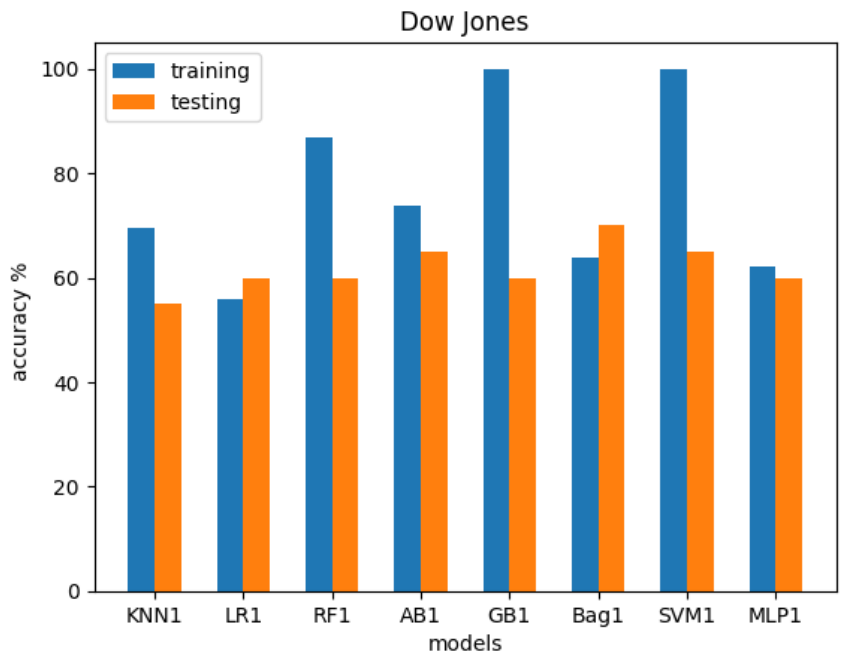

(figure 23: training & testing scores of models on Dow Jones)

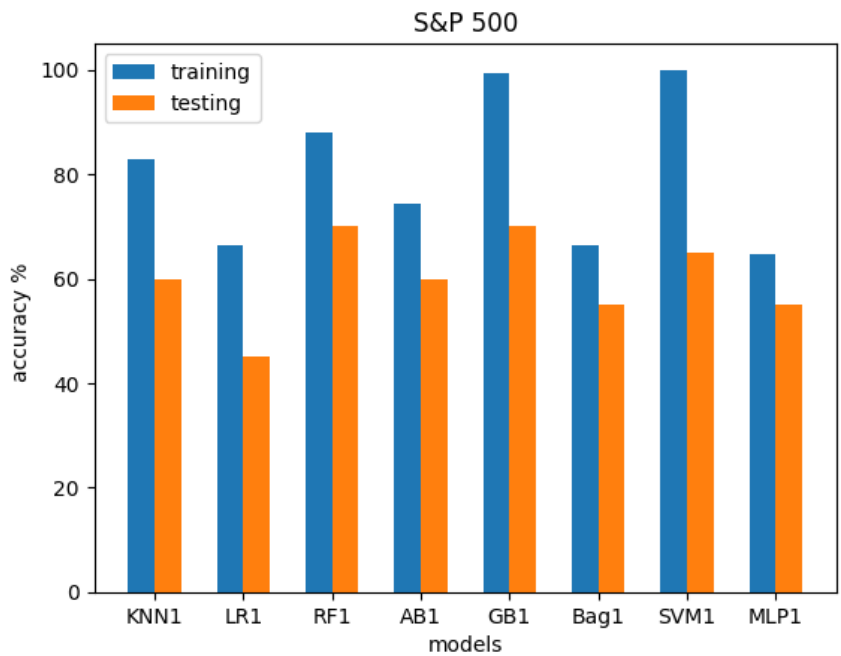

(figure 24: training & testing scores of models on S&P500)

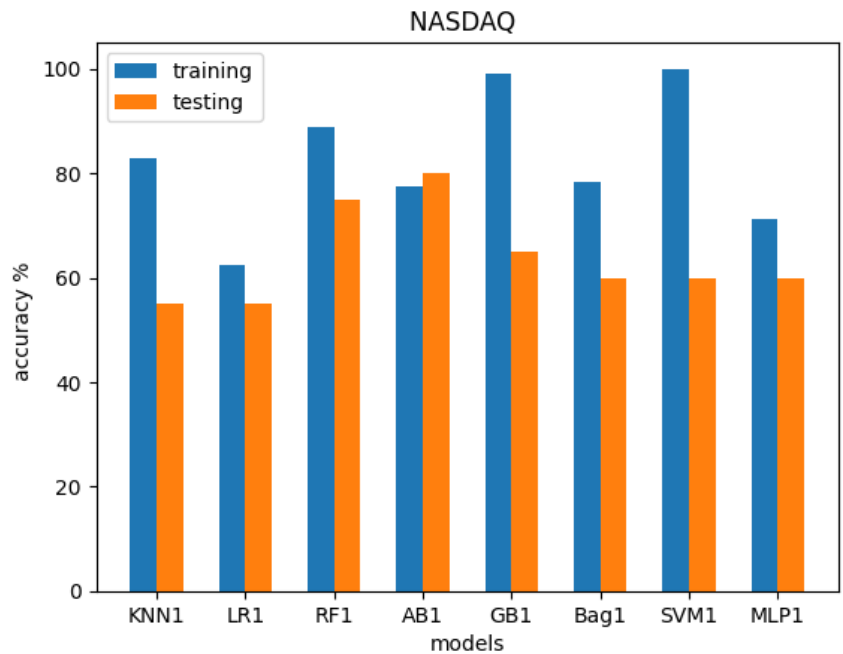

(figure 25: training & testing scores of models on Nasdaq)

(3) The presented chart showcases the financial performance of ML models on prominent American stocks, including Microsoft, Apple, Nvidia, Alphabet, Amazon, and Meta. The benchmark of random traders demonstrates a positive net return on all stocks except for Apple, and outperforms several models across multiple stocks.
   a) In general, the performance of all models is similar for Apple and Microsoft. However, for Nvidia and Meta, the performance of different models exhibits significant variations.
   b) Specifically, SVM, LR and KNN exhibit the lowest performance for Apple, Alphabet and Meta respectively, lagging behind the benchmark by approximately 5%, 15% and 20% respectively.
   c) RF, AB, and GB demonstrate mostly positive returns on all stocks, but performance below the benchmark by 15%-20% on Nvidia. Bagging performs better than the benchmark on most stocks, except for Nvidia, where it has a difference of around 15%.
   d) Only MLP surpasses the baseline performance for each stock, displaying stability and robustness.

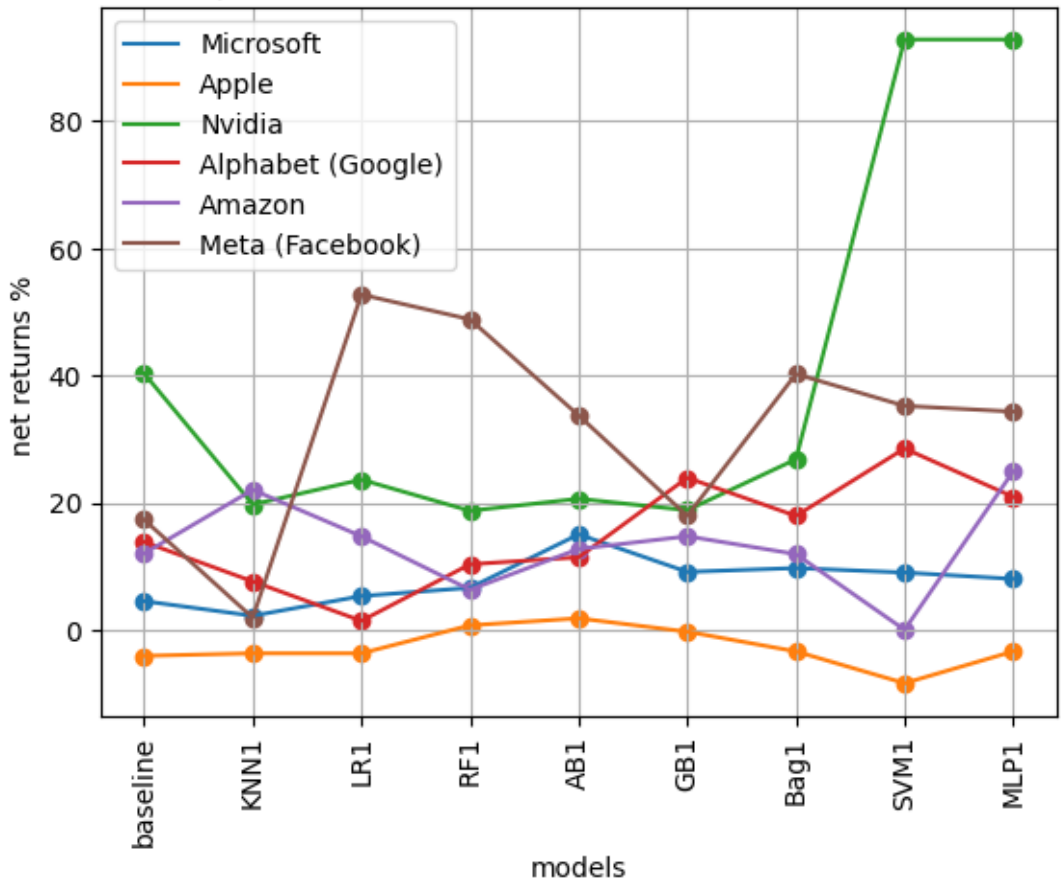

(figure 26: financial performance of different models on top American stocks)
(4) The following charts present the technical performance of ML models on the leading American stocks.
   a) RF, AB, GB, and SVM exhibit signs of overfitting across all stocks, while KNN overfits specifically on Nvidia and underperforms on a few other stocks with accuracies below 50%.
   b) LR underperforms on all stocks except Meta. In contrast, MLP demonstrates the most stable performance, with training and testing accuracies being closer to each other and all testing accuracies equal to or greater than 50%.
   c) MLP is much better and more stable, training and testing accuracies are closer
   d) Also, there are a few instances where the testing accuracy slightly surpasses the training accuracy, which could be attributed to the utilization of data augmentation techniques.

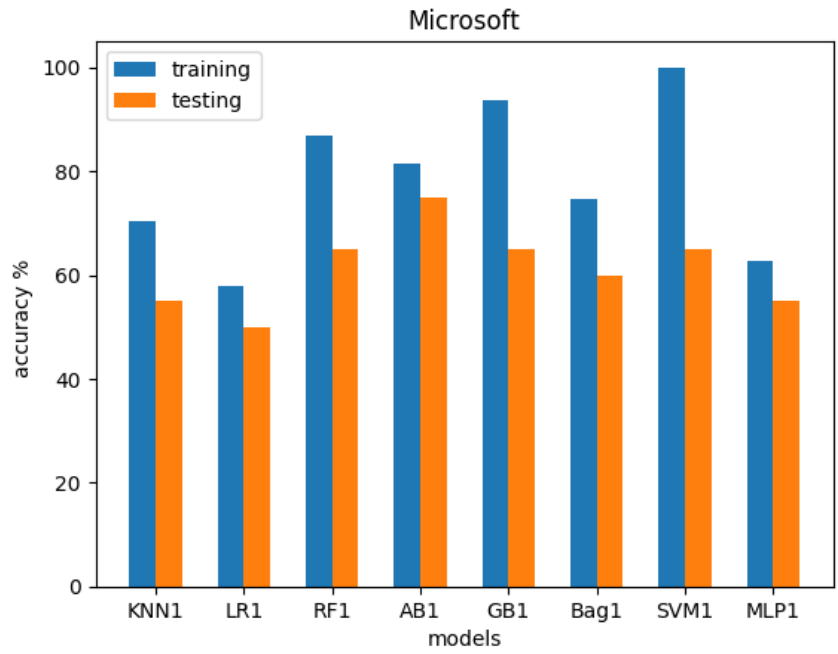

(figure 27: training & testing scores of models on Microsoft)

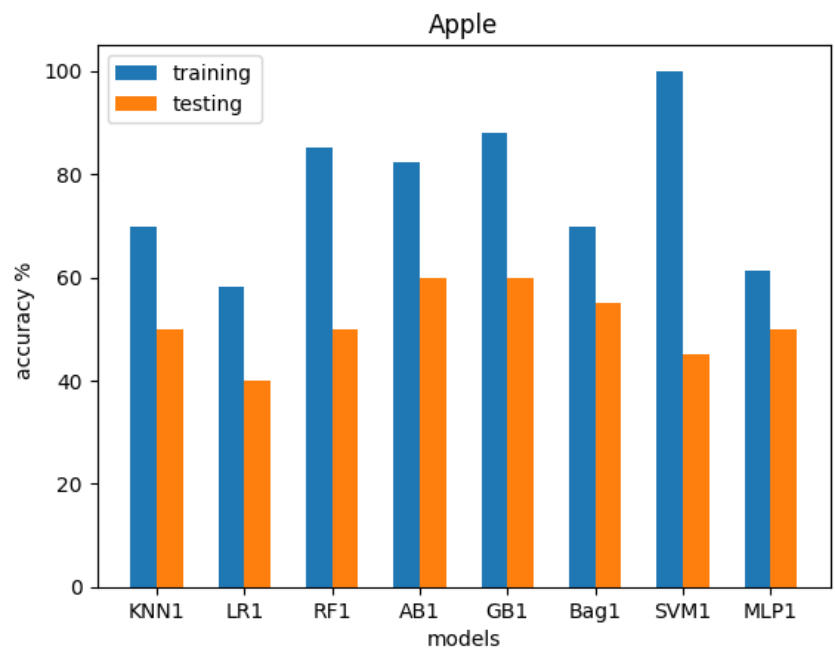

(figure 28: training & testing scores of models on Apple)

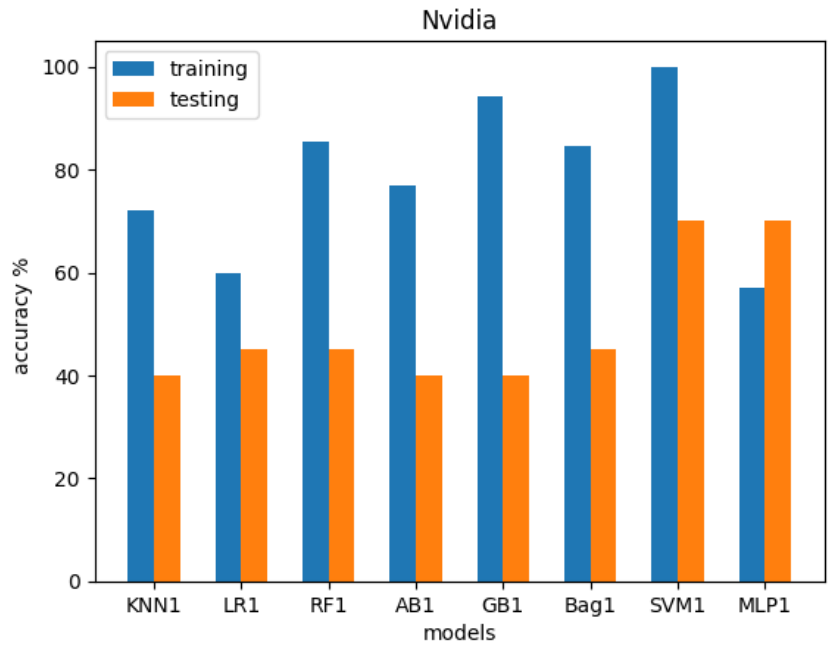

(figure 29: training & testing scores of models on Nvidia)

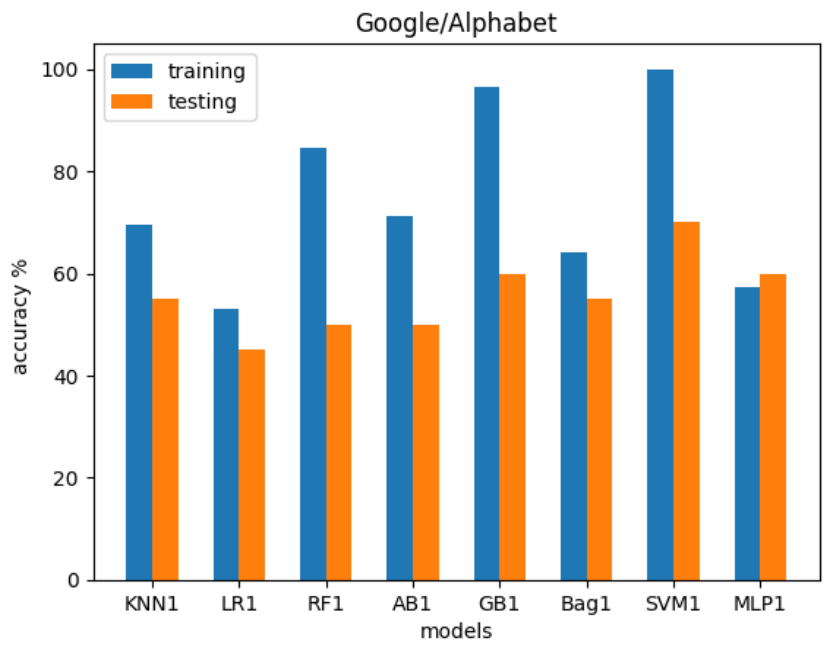

(figure 30: training & testing scores of models on Google/Alphabet)

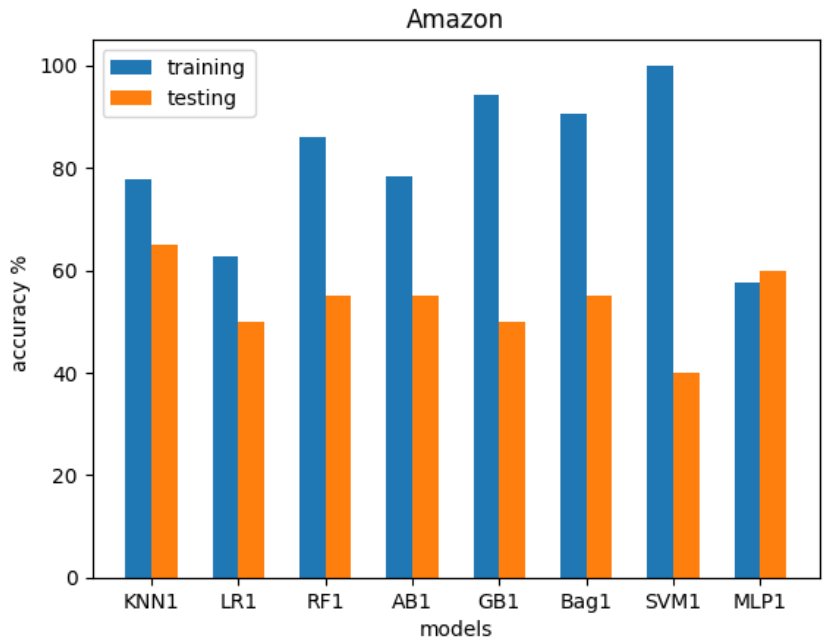

(figure 31: training & testing scores of models on Amazon)

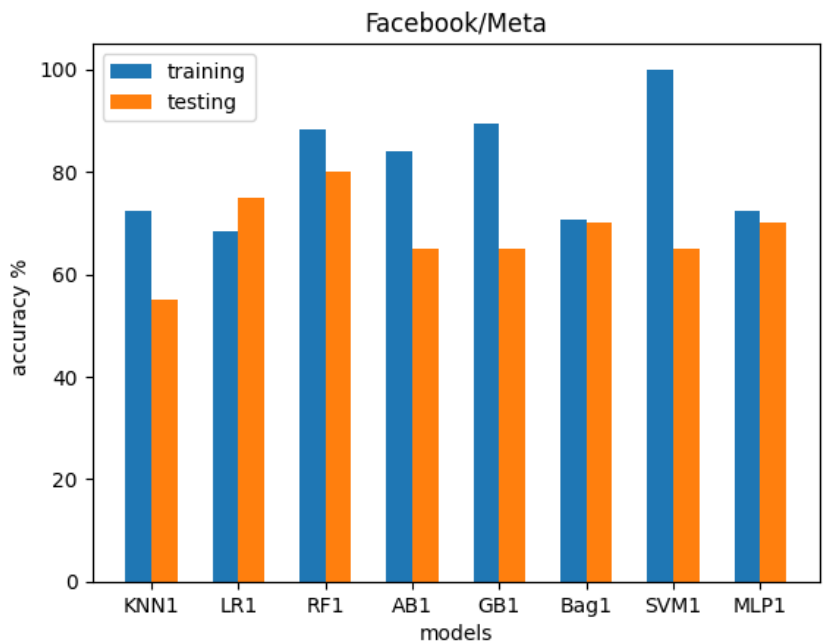

(figure 32: training & testing scores of models on Facebook/Meta)

(5) Overall, MLP outperforms the benchmark of random traders on all indexes and stocks, which demonstrates stability and robustness. Meanwhile, the benchmark itself exhibits a strong performance to a certain extent.

# 5. Conclusion

This study conducts research on the weekly movement prediction of stock markets and shows a clear message that using ML techniques for weekly movement prediction can be both feasible and profitable. This work compares the performance of ML models against objective and independent benchmark, and also provides comprehensive training and testing information on extensive data. This approach enhances the credibility of the findings when compared with several studies that lack transparency in reporting training accuracy or related information.

However, it is important to acknowledge the limitations of this work. Firstly, it does not account for market frictions and market impacts that can influence trading outcomes. Additionally, the study only tests the models on developed markets and does not assess their performance in emerging markets. Furthermore, the work does not consider the short-selling mechanism for trading strategies and lacks a risk analysis pertaining to the trading strategies based on the model's predictions. These limitations present avenues for future research to explore and address these topics in greater depth.